\documentclass[aps,prd,preprint,superscriptaddress,tightenlines,nofootinbib]{revtex4}

\usepackage{graphicx}
\usepackage{dcolumn}
\usepackage{bm}

\usepackage{amsmath}

\newcommand{\Ufs}{$\Upsilon(4\text{S})$}
\newcommand{\Uis}{$\Upsilon(1\text{S})$}
\newcommand{\MUis}{\Upsilon(1\text{S})}
\newcommand{\gev}{\,\rm GeV}
\newcommand{\mev}{\,\rm MeV}

\newcommand{\mom}{\,\rm GeV/c}
\newcommand{\mass}{\,\rm GeV/c^2}
\newcommand{\miss}{\,\rm MeV/c^2}
\newcommand{\fbi}{\,\rm fb^{-1}}

\begin{document}
\preprint{CLNS 05/1929}       
\preprint{CLEO 05-17}         
\title{Radiative Decays of the \Uis \\ to a Pair
of Charged Hadrons}

\author{S.~B.~Athar}
\author{P.~Avery}
\author{L.~Breva-Newell}
\author{R.~Patel}
\author{V.~Potlia}
\author{H.~Stoeck}
\author{J.~Yelton}
\affiliation{University of Florida, Gainesville, Florida 32611}
\author{P.~Rubin}
\affiliation{George Mason University, Fairfax, Virginia 22030}
\author{C.~Cawlfield}
\author{B.~I.~Eisenstein}
\author{G.~D.~Gollin}
\author{I.~Karliner}
\author{D.~Kim}
\author{N.~Lowrey}
\author{P.~Naik}
\author{C.~Sedlack}
\author{M.~Selen}
\author{E.~J.~White}
\author{J.~Williams}
\author{J.~Wiss}
\affiliation{University of Illinois, Urbana-Champaign, Illinois 61801}
% ...
\collaboration{CLEO Collaboration} 
\noaffiliation


\date{October 4, 2005}

\begin{abstract}
Using data obtained with the CLEO~III detector, running at the Cornell
Electron Storage Ring (CESR),
we report on a new study of exclusive
radiative \Uis~decays into the final states
$\gamma\pi^+\pi^-$, $\gamma K^+ K^-$, and $\gamma p \bar{p}$. 
We present branching ratio measurements
for the decay modes $\MUis \rightarrow \gamma f_2(1270)$, $\MUis
\rightarrow \gamma f_2'(1525)$, and $\MUis
\rightarrow \gamma K^{+}K^{-}$; helicity production ratios for
$f_2(1270)$ and $f_2'(1525)$; upper limits for the decay $\MUis
\rightarrow \gamma f_J(2200)$, with $f_J(2220) \rightarrow\pi^+\pi^-$,
$K^+ K^-$, $p \bar{p}$; and an upper limit for the
decay $\MUis\rightarrow \gamma X(1860)$, with $X(1860) \rightarrow
\gamma p \bar{p}$.     
\end{abstract}

\maketitle
\section{Introduction}

Radiative decays of heavy quarkonia, where a photon replaces one of the
three gluons from the strong decay of, for example, the $J/\psi$ or \Uis, are
useful in studying color-singlet two-gluon systems. The two
gluons can, among other things, hadronize into a
meson\footnote{Several authors have studied meson
production in \Uis~radiative decays, giving predictions for
branching and helicity production ratios. The heavy-quarkonium
system is usually described by non-relativistic QCD~\cite{nrqcd}, while the
gluonic hadronization has been treated using soft collinear effective
theory~\cite{ugg}, gluon distribution amplitudes~\cite{nice}, and
perturbative QCD~\cite{russ1,russ2}.},
or directly form a glueball\footnote{Glueballs are a 
natural consequence of QCD, and predictions of their properties
have been made 
using different approaches, such as potential
models~\cite{pot1,pot2,pot3}, lattice QCD 
calculations~\cite{theory1,theory2,lat1,lat2}, bag
models~\cite{bag1,bag2,bag3,theory3},  
flux-tube models~\cite{theory4}, the QCD sum rules~\cite{theory5},
the Bethe-Salpeter (B-S) equation~\cite{bs1,bs2}, QCD
factorization formalism models~\cite{theory6,cleof},
weakly-bound-state models~\cite{maurizio}, and a three-dimensional
relativistic equation~\cite{sre}.  
However,
despite intense experimental
searches~\cite{expglu1,expglu2,expglu3,experiment1,experiment2,experiment3,experiment4}, 
there is 
no conclusive experimental evidence of their direct observation, although
there are strong indications that glueballs contribute to the rich
light
scalar~\cite{noglu1,noglu2,noglu3,noglu4,noglu5,noglu6,noglu7,noglu8,noglu9,noglu10}
and tensor~\cite{tensor1,tensor2} spectrums.}.
Further  
information on radiative decays of heavy quarkonia can be found
in~\cite{QWG}.


Light-meson production in $J/\psi$ two-body radiative decays has been 
experimentally well established with branching fractions at the   
$10^{-3}$ level, based largely on evidence provided by radiative decays
to a pair of 
hadrons\footnote{We refer to the 
Particle Data Group~\cite{pdg} for a
summary of $J/\psi$ radiative decays.}. The production
ratios of the available helicity states have 
been measured for the tensor mesons
$f_2(1270)$~\cite{pluto,markII,crystal,dm2} and
$f_2^{'}(1525)$~\cite{markIII,kkBes}  
in $J/\psi$ two-body radiative decays and 
agree with
theoretical predictions~\cite{very_old,correction2}. In 1996, the BES  
Collaboration reported the observation of the $f_J(2220)$ in
$J/\psi$ two-body radiative decays, and measured product branching
fractions, ${\cal B}(J/\psi \rightarrow \gamma f_J(2220))\times{\cal
B}(f_J(2220)\rightarrow h^+h^-)$ (we use the convention $h = \pi,\ K,\ p$), of
the order of $10^{-5}$~\cite{glueball}. Much excitement 
was generated at the time 
because it is possible to interpret the $f_J(2220)$ as a
glueball. A candidate similar to $f_J(2220)$ was reported in 1986 by
the {Mark III} 
collaboration in the $K\bar{K}$ mode~\cite{markIIIglue}, but was not
confirmed by the DM2 
collaboration~\cite{dm2glue}. Recently, BES reported the existence
of a new particle, the $X(1860)$, observed by its decay 
$J/\psi \rightarrow \gamma X(1860) \rightarrow p
\bar{p}$~\cite{ppbar}, a result that is 
currently being interpreted~\cite{pp1,pp2,pp3,pp4,pp5,pp6,pp7}.


The experimental observation of radiative \Uis~decays is challenging 
because their rate is suppressed to a level of
\begin{equation*}
\left (\frac{q_b}{q_c}\right )^2\left (\frac{m_c}{m_b}\right )^2 \approx 0.025,
\end{equation*}
of the corresponding rate of $J/\psi$ 
radiative decays. This factor arises because the quark-photon
coupling is proportional to the electric charge, and the quark
propagator is roughly proportional to $1/m$ for low momentum
quarks. Taking into account the
total widths~\cite{pdg} of $J/\psi$ and \Uis, the branching
fraction of a particular \Uis~radiative decay mode is expected to be
around $0.04$ of the corresponding $J\psi$ branching fraction. 
In 1999, CLEO~II made the first 
observation of a radiative \Uis~decay to a pair of hadrons~\cite{cleo1}, 
which was consistent with $\MUis \rightarrow \gamma
f_2(1270)$, where $f_2(1270) \rightarrow
\pi\pi$. Comparing the measured branching fraction to 
the $J/\psi \rightarrow \gamma f_2(1270)$ branching fraction, a suppression
factor of $0.06 \pm 0.03$ was obtained. Recent theoretical
work~\cite{ugg,nice} predict
a suppression 
factor between $0.06-0.18$ for this mode, and favor the production of
$f_2(1270)$ in a helicity-0 state. After the BES result for the
$f_J(2220)$ in radiative $J/\psi$ decays, a corresponding search was performed
by CLEO~II in the radiative \Uis~system~\cite{cleo2} and
limits were put on some of the glueball candidates' product branching ratios. 

In this paper, we use the CLEO~III \Uis~data sample, which has
fifteen times higher statistics and better particle 
identification than the CLEO~II data sample, to probe the color-singlet
two-gluon spectrum by measuring the system's invariant mass using its
decays to $\pi^+\pi^-$, $K^+K^-$, and $p \bar{p}$.  
Further details of this analysis can be found elsewhere~\cite{LUIS}.

\section{CLEO~III Detector, Data, and Monte Carlo Simulated Sample}

The CLEO~III detector is a versatile multi-purpose particle
detector described more fully in~\cite{cleoiii}. It is centered on
the interaction region of CESR. From the $e^+e^-$   
interaction region radially outward it consists of a silicon strip
vertex detector and a wire drift chamber used to measure the position, momenta,
and specific ionization energy losses ($dE/dx$)
of charged tracks based on their fitted path in a 1.5\,T solenoidal
magnetic field and the amount of charge
deposited on the drift chamber wires. The silicon vertex detector and
drift chamber tracking system achieves a
charged particle momentum resolution of 0.35\%(1\%) at 1\mom(5\mom)
and a $dE/dx$ resolution of 6\%.
Beyond the drift chamber is a Ring Imaging Cherenkov
Detector, RICH, which covers 80\% of the solid angle and is
used to further identify charged particles by giving for each mass 
hypothesis the likelihood of a fit to the Cherenkov radiation
pattern. After the RICH is a Crystal  
Calorimeter (CC) that covers 93\% of the solid angle. The CC has a
resolution of 2.2\% (1.5\%) for 1\gev (5\gev) photons. After the CC is
a superconducting solenoid coil that provides the magnetic field,
followed by iron flux return plates 
with wire chambers interspersed in three layers at 3, 5, and 7 hadronic
interaction lengths to provide muon identification.

The data sample has an integrated luminosity of $1.13\fbi$ taken at
the \Uis~energy, $\sqrt{s}=9.46\gev$, which 
correspond to $21.2\pm0.2$ million \Uis~decays~\cite{nuis} and $3.49\fbi$ taken
at the \Ufs~energy, $\sqrt{s}=10.56\gev$, used to model the underlying
continuum present in 
the \Uis~data sample. The continuum background modeling is important because  
continuum background processes such as $e^+e^- \rightarrow \gamma
\rho$ with $\rho \rightarrow  \pi^+ \pi^-$, $e^+e^- \rightarrow \gamma
\phi$ with $\phi \rightarrow  K^+ K^-$, and direct $e^+e^- \rightarrow \gamma
h^+ h^-$ have the same
topology as the signal events we are investigating. 

Efficiencies are evaluated using a Monte Carlo simulation of the process~\cite{qq}
and a GEANT-based~\cite{geant} detector response. Monte Carlo samples of
$e^+e^- \rightarrow \gamma X$ with $X \rightarrow h^+ h^-$ are generated at both the
\Uis~and \Ufs~energies with uniform angular distributions and flat $h^+
h^-$ invariant mass distributions from threshold to $3.5\mass$.

\section{Event Selection}

Events which satisfy the CLEO~III trigger~\cite{trigger} are then required to
meet the following analysis requirements: (a) There are exactly two charged
tracks that trace back to the beamspot and have good quality
track fits and $dE/dx$
information.
(b) There is exactly one 
CC shower that is unmatched to any
track and whose energy, $E_{\gamma}$, is greater than $4\gev$. 

Each event is also required
to be consistent with having the 4-momentum of the initial $e^+e^-$
system by demanding that the
chi-squared from a kinematic fit to the following constraint,  
\begin{equation}
	\vec{p}_{h^+h^-} + (2E_{\rm beam}-E_{h^+h^-})\widehat{p}_{\gamma} 
	= \vec{p}_{CM},\label{constraint}
\end{equation}
be less than 100, where $\vec{p}_{h^+h^-}$ is the di-hadron momentum,
$E_{h^+h^-}$ is 
the di-hadron energy, $E_{\rm beam}$ is the beam energy,
$\widehat{p}_{\gamma}$ is the photon's direction, and
$\vec{p}_{CM}$ is the momentum of the $e^+e^-$ system (which has a
magnitude of a few
MeV/c because of the small ($\approx 2$ mrad) crossing angle of the $e^+$ and
$e^-$ beams).
Equation~\ref{constraint} is a  
3-constraint subset of the 4-momentum 
constraint and has the convenient property of avoiding the use of the
measured photon energy, which has an asymmetric
measurement uncertainty. We improve the measurement of the di-hadron
4-momenta (the di-hadron invariant mass resolution becomes 3.2, 2.6,
and 2.0$\miss$ for the pion, kaon, and proton modes, respectively)
by using the constraint in Equation~\ref{constraint}, and then demanding that:
\begin{equation*}
0.950 <  (E_{h^+h^-} + E_\gamma)/2E_{\rm beam} < 1.025.
\end{equation*} 

Strong electron and muon vetoes are imposed to
suppress the abundant QED processes $e^+e^- \rightarrow \gamma e^+ e^-$ and $e^+e^- \rightarrow
\gamma \mu^+ \mu^-$. To reject $e^+e^- \rightarrow \gamma e^+e^-$, we
require 
each track to have a matched CC shower with an energy $E$, together with a measured
momentum $p$, such that
$|E/p-0.95| > 0.1$, and that the combined RICH and $dE/dx$ likelihood
for $h$ be higher than the combined likelihood for $e$. To 
reject $e^+e^- \rightarrow \gamma \mu^+\mu^-$, we require that neither track
produce a signal in
the five hadronic interaction lengths of the muon system. For the
$\pi^+\pi^-$ mode, where muon background is a particular problem because
of the similar pion and muon masses, we further require
that both tracks must be within the
barrel part of the muon chambers ($|\cos\theta| < 0.7$), and both have $p
> 1\mom$. To increase the solid-angle acceptance of the detector and
improve the overall muon suppression efficiency with 
virtually no increase in muon fakes, we flag an event as ``not
muonic'' and remove the muon suppression requirements if either
track deposits more than 600\mev~in the 
CC. 


Events that satisfy all the above requirements
are then identified as either $\pi^+\pi^-$, $K^+K^-$, or $p\bar{p}$
using the RICH and $dE/dx$
information. Since the ratios $\pi^+\pi^-/K^+K^-$ and
$K^+K^-/p\bar{p}$ are much larger than 1 for these types of events, in
the 3 cases in which we try to reduce the background from a lower-mass
hadron, we also use the chi-squared value from
the kinematic constraint in Equation~\ref{constraint} to identify the
event type. Since the constraint involves the di-hadron energy, the
chi-squared value is sensitive to the hadronic masses. After these
procedures, the particle identification efficiencies (fake rates) are 90\% (0.31\%),
99\% (0.03\%), 98\% (0.10\%) for kaons (pions faking kaons), protons (pions faking protons), and
protons (kaons faking protons), respectively.  


\section{Determination of Signals and Their Spin Assignments} 

The overall reconstruction efficiencies as determined by Monte Carlo
simulations, including both event selection and analysis cuts, are
43\%, 48\%, and 56\% for the \Uis~radiative decays to $\pi^+\pi^-$,
$K^+K^-$, and $p\bar{p}$, respectively. These efficiencies are only
mildly dependent on the di-hadron invariant mass and are very similar
for the continuum background events.
The continuum-subtracted di-hadron invariant mass plots are obtained
by efficiency 
correcting each bin of the di-hadron invariant mass plots for the
\Uis~and $\sqrt{s}=10.56\gev$ datasets, 
scaling the latter plot by a factor of 
$0.404\pm0.002$\footnote{We obtain this factor, $f$, from the
integrated luminosities of
the \Uis~and $\sqrt{s}=10.56$\gev~datasets, and the assumption that, to
first order, the cross sections of the continuum processes in each run
are proportional to $1/s$.  This factor is roughly equal to the factor
obtained by using the average energy of each dataset,
\begin{equation*}
f = 0.404 \approx \frac{1.13\fbi}{3.49\fbi} \left ( \frac{10.56\gev}{9.46\gev} \right )^2.
\end{equation*}
}, and subtracting it from the \Uis~dataset invariant mass plot. Possible 
signals are determined by fitting each spectrum to spin-dependent
relativistic Breit-Wigner functions\footnote{The spin-dependent
relativistic Breit-Wigner parameterization used has the following probability
distribution for a particular $h^+h^-$ invariant mass $ x > x_0$,
\begin{equation*}
dP(x) \propto \frac{xx_m\Gamma(x)}{(x^2-x_m^2)^2+(x_m\Gamma(x))^2} dx, 
\end{equation*}
where
\begin{equation*}
\Gamma(x) =
\Gamma_0\left(\frac{x-x_0}{x_m-x_0}\right)^{2S+1}\frac{2(x_m-x_0)^2}{(x-x_0)^2+(x_m-x_0)^2}.
\end{equation*}
In the above expression, $x_m$ and $\Gamma_0$ represent respectively the most
likely mass and 
width, and are allowed to float during the fit. The values of $x_0$ and $S$
are fixed during the fit to the invariant mass threshold for the particular mode and
the spin of the resonance, respectively. The number of events for each
fitted signal candidate is obtained by integrating this Breit-Wigner
parameterization between threshold and $3\ \mass$.}. The spin value for
each Breit-Wigner is 
surmised by identifying each possible resonance in the 
invariant mass plot based on its approximate mass and width. Later,
we confirm these spin assignments for the significant resonances by
inspecting the angular distributions of the \Uis~decay products.

\begin{figure}
\includegraphics*[width=4.0in]{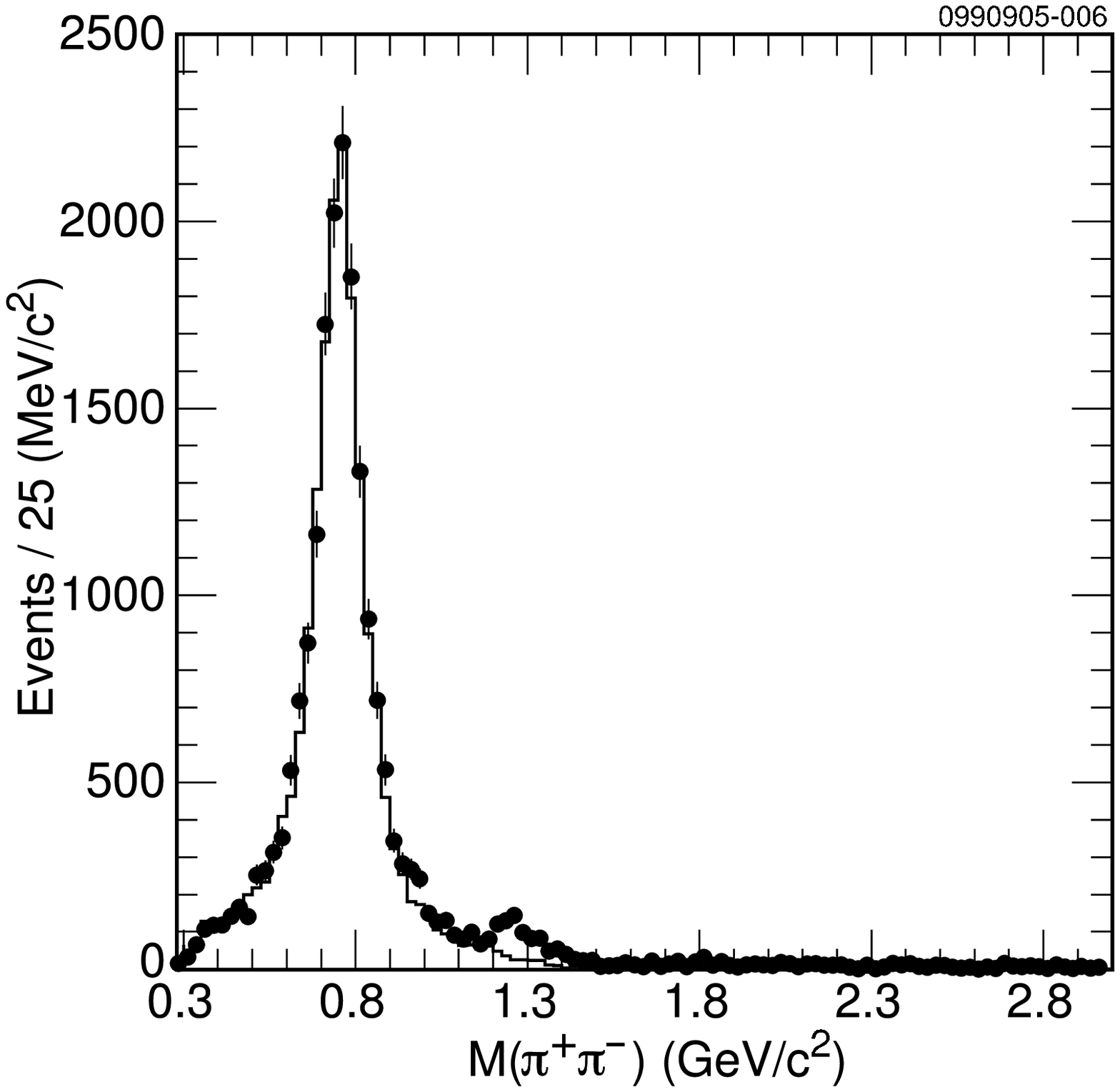}
\caption{Invariant mass of $\pi^+ \pi^-$ from $e^+e^- \rightarrow
\gamma \pi^+ \pi^-$ for the scaled $\sqrt{s}=10.56\gev$ dataset
(solid line), and the \Uis~dataset (circles). The large number of events
near $770\miss$ is due to the abundant process $e^+e^- \rightarrow
\gamma \rho$.} 
\label{paper1.1.eps}
\end{figure}

\begin{figure}
\includegraphics*[width=3.25in]{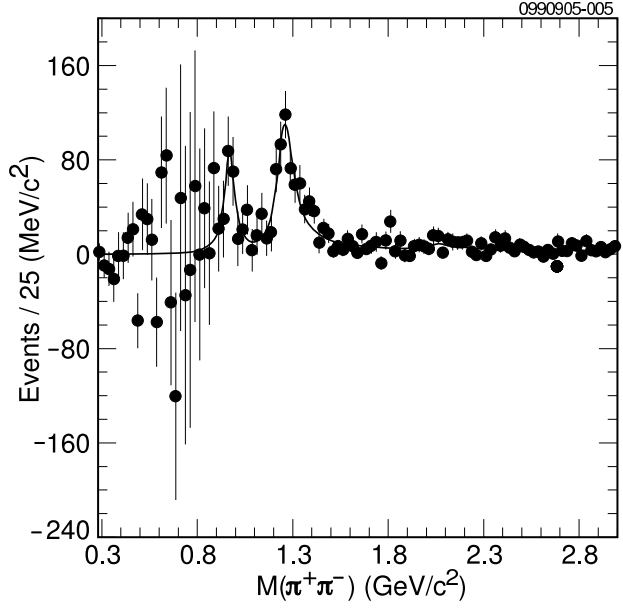}
\caption{Invariant mass of $\pi^+ \pi^-$ from $\MUis
\rightarrow \gamma \pi^+\pi^-$ and the fit to the three spin-dependent
relativistic Breit-Wigner functions described in the text.}
\label{paper1.eps}
\end{figure}

\begin{figure}
\includegraphics*[width=3.25in]{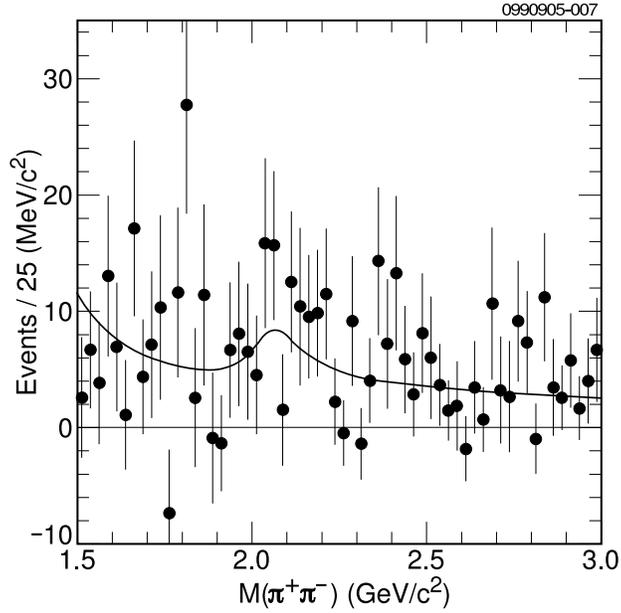}
\caption{Invariant mass of $\pi^+ \pi^-$ from $\MUis
\rightarrow \gamma \pi^+\pi^-$  in the region 1.5-3.0$\mass$. This fit shows
the small, non-significant, excess found in the $f_4(2050)$ region.}
\label{paper1.2.eps}
\end{figure}

The $\pi^+\pi^-$ invariant mass plots for the \Uis~and the scaled
$\sqrt{s}=10.56\gev$ datasets are shown in 
Figure~\ref{paper1.1.eps}. The fit 
to the continuum-subtracted $\pi^+\pi^-$ spectrum, shown in
Figure~\ref{paper1.eps}, has a 
significant $f_2(1270)$ 
signal of $944\pm74$ 
events. It also has   
two less significant signal candidates; $340^{+140}_{-130}$ events
in the $f_0(980)$ region, and $80\pm30$ 
events in the $f_4(2050)$
region (see Figure~\ref{paper1.2.eps}) whose significances are
$4.3\sigma$ and $2.6\sigma$, 
respectively. Each significance is obtained 
by doing multiple chi-squared fits to the invariant mass plot fixing
the signal area to different values, assigning each of these multiple
fits a probability proportional to $e^{-\chi^2/2}$, normalizing the
resulting probability distribution, and calculating the probability for
negative or 0 signal. 

\begin{figure}
\includegraphics*[width=4.0in]{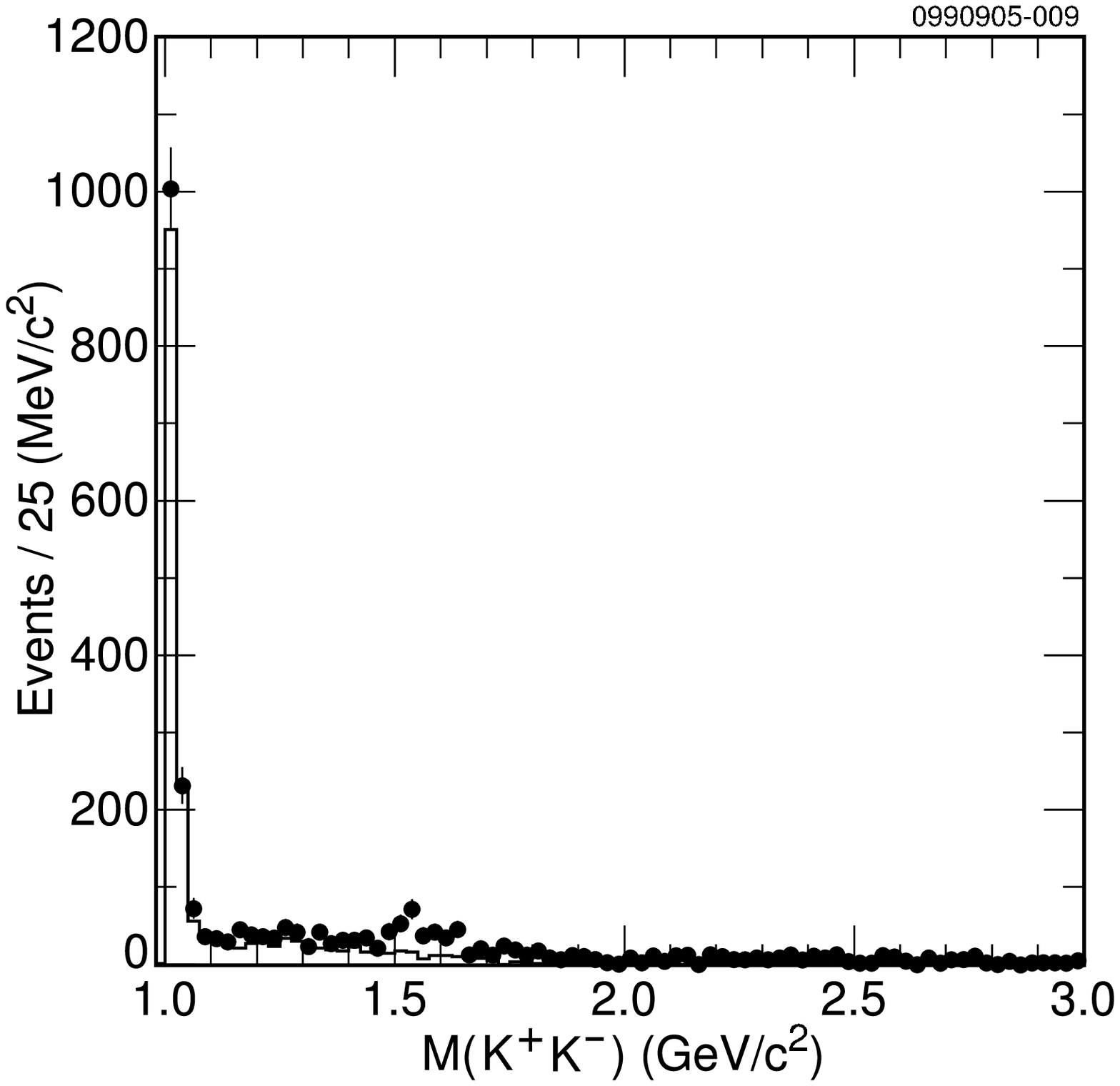}
\caption{Invariant mass of $K^+ K^-$ from $e^+e^- \rightarrow
\gamma K^+ K^-$ for the scaled $\sqrt{s}=10.56\gev$ dataset
(solid line), and the \Uis~dataset (circles). The large number of events
near $1.050\mass$ is due to the abundant process $e^+e^- \rightarrow
\gamma \phi$.} 
\label{paper2.1.eps}
\end{figure}

\begin{figure}
\includegraphics*[width=4.0in]{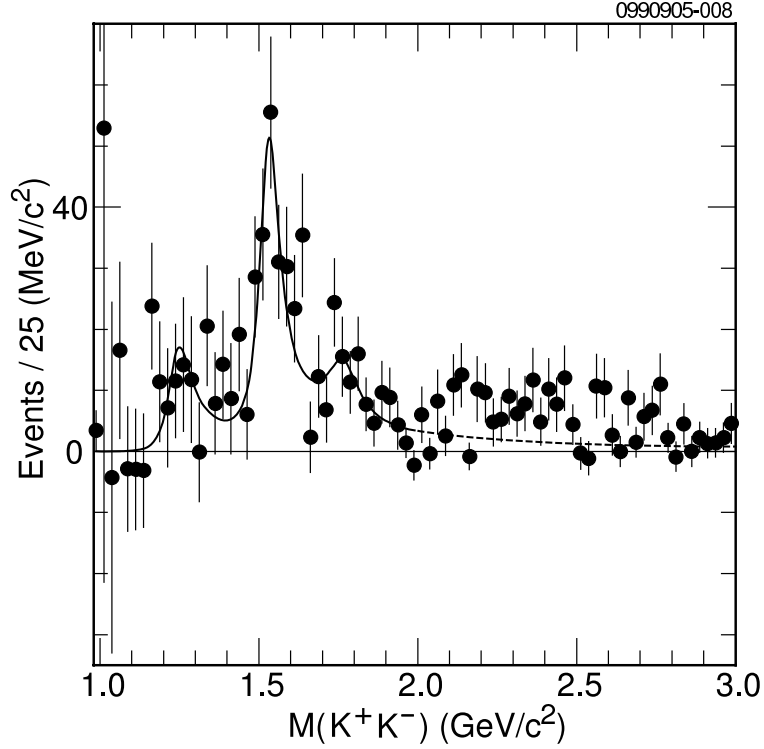}
\caption{Invariant mass of $K^+ K^-$ from $\MUis
\rightarrow \gamma K^+K^-$ and the fit to the three spin-dependent
relativistic Breit-Wigner functions described in the text. The dotted line 
shows the extrapolation of this fit to masses above $2\ \mass$, which is the cut-off
for the fit.}
\label{paper2.eps}
\end{figure}

The $K^+K^-$ invariant mass plots for the \Uis~and the scaled
$\sqrt{s}=10.56\gev$ datasets are shown in 
Figure~\ref{paper2.1.eps}.
The fit to the continuum-subtracted $K^+K^-$
spectrum, shown in Figure~\ref{paper2.eps}, has a 
significant signal of $312^{+69}_{-61}$ 
events identified 
as the $f_2^{'}(1525)$, and two non-significant signal candidates indicating 
possible $f_2(1270)$
and
$f_0(1710)$ production with $109\pm36$ and $73\pm29$ events whose
significances are 
$3.2\sigma$ and  
$3.3\sigma$, respectively. The excess of events in the $f_2(1270)$
region is consistent with that expected 
using the $\gamma\pi^+\pi^-$ data and the known branching ratios for the
$f_2(1270)$. We also note
that there is a significant excess of $220\pm20$ events above
$2.0\ \mass$ in the $K^+K^-$ invariant mass distribution which is not 
associated with any resonant structure.

\begin{figure}
\includegraphics*[width=3.25in]{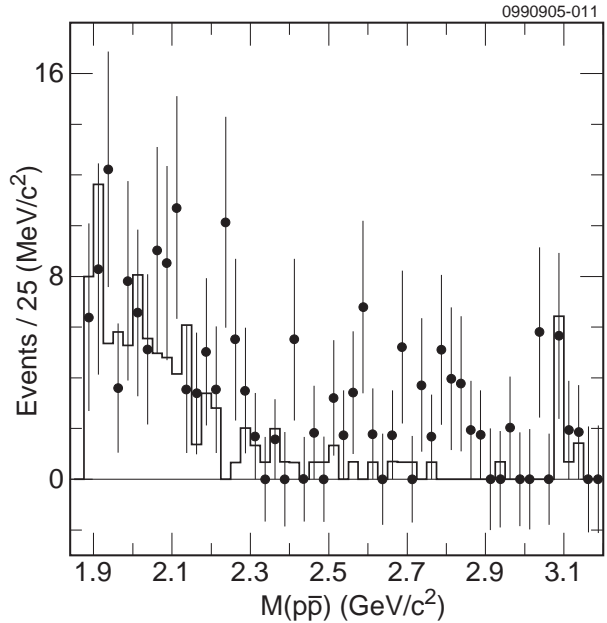}
\caption{Invariant mass of $p \bar{p}$ from $e^+e^- \rightarrow
\gamma p \bar{p}$ for the scaled $\sqrt{s}=10.56\gev$ dataset
(solid line), and the \Uis~dataset (circles). The events
near $3.1\mass$ are due to the process $e^+e^- \rightarrow \gamma J/\psi$.}
\label{paper3.1.eps}
\end{figure}

\begin{figure}
\includegraphics*[width=3.25in]{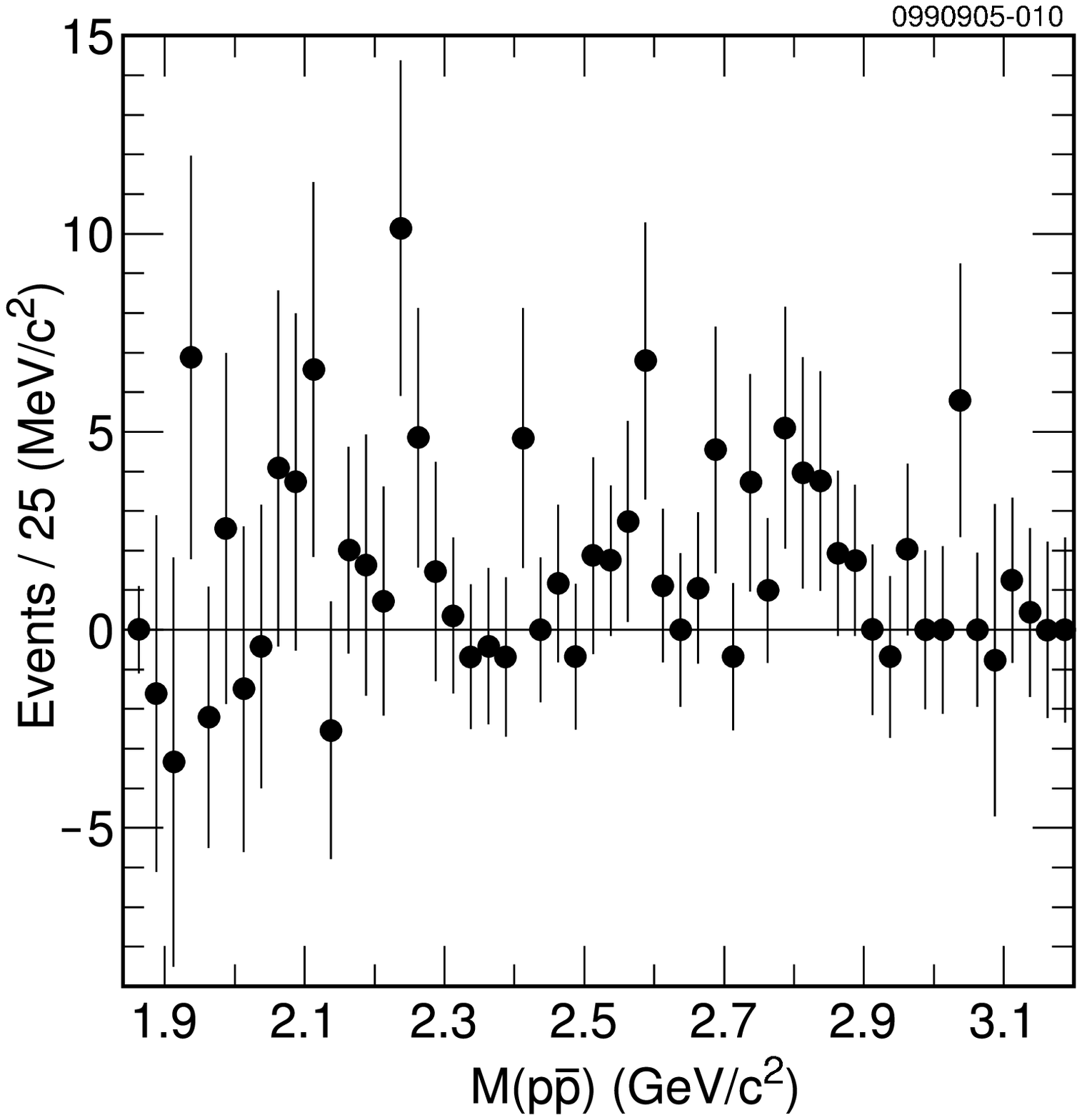}
\caption{Invariant mass of $p \bar{p}$ from $\MUis
\rightarrow \gamma p \bar{p}$.}
\label{paper3.eps}
\end{figure}

The $p\bar{p}$ invariant mass plots for the \Uis~and the scaled
$\sqrt{s}=10.56\gev$ datasets are shown in 
Figure~\ref{paper3.1.eps}.
No recognizable structure is seen in the continuum-subtracted $p\bar{p}$ 
spectrum, which is shown in Figure~\ref{paper3.eps}. In particular, we do
not see an enhancement near threshold, as might be expected from the BES $X(1860)$
results \cite{ppbar}. There is a non-significant  excess
of $85\pm18$ events in the $2-3\mass$ invariant mass region. 


\begin{figure}
\includegraphics*[width=3.25in]{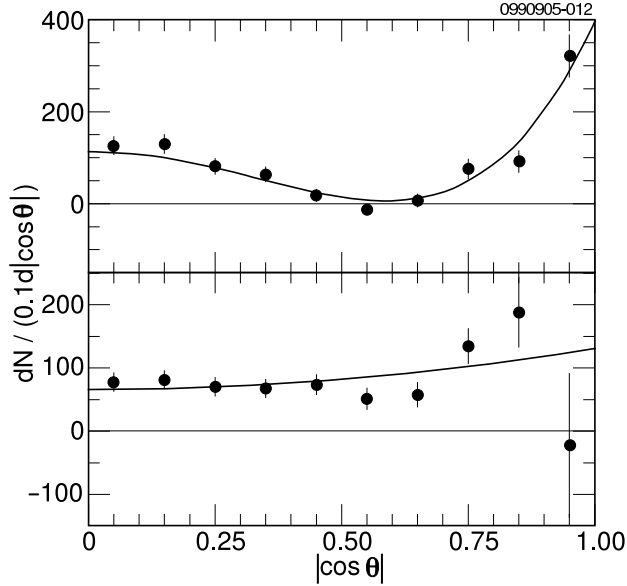}
\caption{Distributions of $|\cos\theta_{\pi}|$ (top) and
$|\cos\theta_{\gamma}|$ (bottom) for the signal events in the
$f_2(1270)$ invariant mass region. The solid lines correspond to a
simultaneous fit to the $J = 2$ helicity formalism prediction
(Equation~\ref{simul}).} 
\label{paper4.eps}
\end{figure}

\begin{figure}
\includegraphics*[width=3.25in]{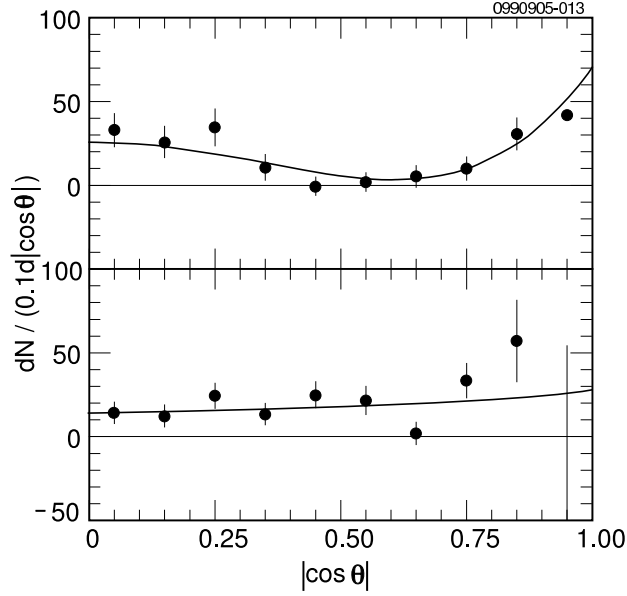}
\caption{Distributions of $|\cos\theta_{K}|$ (top) and
$|\cos\theta_{\gamma}|$ (bottom) for the signal events in the
$f_2^{'}(1525)$ invariant mass region. The solid lines correspond to a
simultaneous fit to the $J = 2$ helicity formalism prediction
(Equation~\ref{simul}).} 
\label{paper5.eps}
\end{figure}

To confirm the spins of our $f_2(1270)\rightarrow \pi^+\pi^-$ and 
$f_2^{'}(1525)\rightarrow K^+K^-$ 
signals, we
examine the absolute value of the cosine of the polar angle of the photon
with respect to 
the beam axis, 
$|\cos\theta_{\gamma}|$, and the absolute value of the cosine of the
angle formed by the 
3-momentum vector of one of the hadrons measured in the di-hadron rest
frame with 
the photon's direction,
$|\cos\theta_{h}|$. The event selection 
efficiency is slightly dependent on both angles, so to minimize
systematic effects, the $|\cos\theta_{\gamma}|$ and
$|\cos\theta_{h}|$ efficiency-corrected distributions are obtained by
projecting the 
2-dimensional bin-by-bin efficiency-corrected
$(|\cos\theta_{\gamma}|,|\cos\theta_{h}|)$ 
distribution. We also subtract the background contributions from the tails
of nearby resonances. 
The resulting angular distributions (shown in Figures~\ref{paper4.eps}
and~\ref{paper5.eps}) are simultaneously fit to the helicity formalism
prediction~\cite{correction2,LUIS,helicity} for
different resonance spin hypotheses up 
to $J = 4$. For the $f_2(1270)$ the different fit confidence levels
are $8\times10^{-19}$, $2\times10^{-19}$, $0.05$, $8\times10^{-12}$, and
$1\times10^{-12}$ for the hypotheses $J = 0,1,2,3,4$,
respectively.  For the $f_2^{'}(1525)$ the different fit confidence levels
are $2\times10^{-4}$, $2\times10^{-4}$, $0.23$, $8\times10^{-3}$, and
$2\times10^{-3}$ for the hypotheses $J =
0,1,2,3,4$, respectively. These results confirm our identification of the
resonances, as in both cases the angular distributions of the data
strongly favor the $J = 2$ 
hypothesis\footnote{Some
authors use a 
probability distribution that also depends on a third angle,
$\phi_h$~\cite{markIII}. However, extreme care must be taken when
using this angle 
because it makes the probability distribution sensitive to the
relative phases of 
the helicity amplitudes. Thus, two new free parameters need to be introduced
in such a probability distribution, as was noted by~\cite{correction2}
and correctly implemented 
by~\cite{markIII,dm2}. Otherwise, the measurement of the helicity amplitudes
rests on the assumption that their relative phases are
0~\cite{old_embarrasment,pluto,crystal,embarrasment}.},  
\begin{equation}
\begin{split}
\frac{dP_{\theta_h,\theta_{\gamma}}}{d\cos\theta_hd\cos\theta_{\gamma}} =
&|a_0|^2\times\frac{5}{8}(3\cos^2\theta_h-1)^2
\times\frac{3}{8}(1+\cos^2\theta_{\gamma})+ \\ 
&|a_1|^2\times\frac{15}{16}\sin^22\theta_h
\times\frac{3}{4}\sin^2\theta_{\gamma}+\\ 
&|a_2|^2\times\frac{15}{16}\sin^4\theta_h
\times\frac{3}{8}(1+\cos^2\theta_{\gamma}), 
\end{split}\label{helicity}
\end{equation}
where $a_{\lambda},\ \lambda=0,\ 1,\ 2,$ are the normalized helicity
amplitudes, $\int dP_{\theta_h,\theta_{\gamma}} =
|a_0|^2+|a_1|^2+|a_2|^2 = 1$. In other words, 
$|a_{\lambda}|^2$ is the  
probability of $X$ in $\MUis \rightarrow 
\gamma X$ to have helicity $\pm\lambda$.  Because of the normalization
condition, the $(\theta_h,\theta_{\gamma})$ probability
distribution can be described by two free parameters, 
traditionally chosen to be the helicity production ratios, 
\begin{equation*}
\begin{split}
x^2 = 
\frac{|a_1|^2}{|a_0|^2} \text{ and } 
y^2 = 
\frac{|a_2|^2}{|a_0|^2}.
\end{split}
\end{equation*}
To measure $x^2$ and $y^2$, we simultaneously fit the
data to the individual $\theta_h$ and 
$\theta_{\gamma}$ distributions using\footnote{We choose to use a
simultaneous fit to these two distributions instead 
of a two-dimensional fit using Equation~\ref{helicity}
because of our limited statistics.}  
\begin{equation}
\begin{split}
\frac{dN_{\theta_{\gamma}}}{d\cos\theta_{\gamma}} &= 
N\int_{\theta_h}dP_{\theta_h,\theta_{\gamma}} \\ &=
\frac{N}{1+x^2+y^2}
\left [
\frac{3}{8}(1+y^2)(1+\cos^2\theta_{\gamma})
+\frac{3}{4}x^2\sin^2\theta_h
\right ]
\\ 
\frac{dN_{\theta_h}}{d\cos\theta_h} &= 
N\int_{\theta_{\gamma}}dP_{\theta_h,\theta_{\gamma}} \\ &=
\frac{N}{1+x^2+y^2}
\left [\frac{5}{8}(3\cos^2\theta_h-1)^2+
\frac{15}{16}x^2\sin^22\theta_h+
\frac{15}{16}y^2\sin^4\theta_h \right ], 
\end{split}\label{simul}
\end{equation}
where $N$ corresponds to the number of events. Using the fits to the
data (see Figures~\ref{paper4.eps} and~\ref{paper5.eps}) we measure the
following helicity production ratios:
\begin{equation*}
\begin{split}
&x^2_{f_2(1270)} = 0.00^{+0.02+0.01}_{-0.00-0.00}\ \ \ \
y^2_{f_2(1270)} = 0.09^{+0.08+0.04}_{-0.07-0.03}\\
\\
&x^2_{f_2^{'}(1525)} = 0.00^{+0.10+0.01}_{-0.00-0.00}\ \ \ \
y^2_{f_2^{'}(1525)} = 0.30^{+0.22+0.07}_{-0.17-0.06}\\
\end{split}
\end{equation*}
where the first uncertainty is statistical and the second is
systematic. The systematic uncertainty is quantified by studying
how well input values are reproduced when analyzing Monte Carlo 
samples,
the measured angular
distribution of the photon and 
tracks from $e^+e^-\rightarrow \gamma \rho, \gamma \phi$ in 
the $\sqrt{s} = 10.56$ dataset, the effect of a one-sigma variation in
the continuum scale factor, and possible interference with nearby
resonances. We find that possible interference with nearby resonances in the
$|\cos\theta_{h}|$ distribution
dominates the systematic uncertainty. 
The helicity production ratio measurements indicate that both  
resonances are   
predominantly produced with helicity 0. They are in agreement with
the predictions of~\cite{ugg}, and in good agreement with the
twist-two-order predictions of~\cite{nice}:
no $\lambda = 1$ production, 
and $\lambda = 2$ production suppressed by a factor of
$(m_{X}/m_{b})^2$ with respect to $\lambda = 0$ production, where
$m_{X}$ is the mass of the tensor meson and $m_{b}$ is the mass
of the $b$ quark.   




We use the results from fitting the angular distributions to
correct the Monte Carlo simulation efficiencies, which are calculated
using flat distributions in the relevant angles, by a 
factor of $0.78\pm0.02$ for the $f_2(1270)$, $0.90\pm0.01$ for the
$f_2^{'}(1525)$, and $0.88^{+0.03}_{-0.01}$ for the significant excess in
the $2-3\mass$ region of the di-kaon invariant mass. 
The large correction in the pion mode is due to the
necessarily stronger muon suppression requirement. The
measured branching ratios of the significant resonances are:
\begin{equation*}
\begin{split}
&{\cal B}(\MUis \rightarrow \gamma f_2(1270)) =
(10.2 \pm0.8\pm0.7)\times10^{-5}, \\
&{\cal B}(\MUis \rightarrow \gamma f_2^{'}(1525)) = 
(3.7^{+0.9}_{-0.7}\pm0.8)\times10^{-5},
\end{split}
\end{equation*}
and the measured branching ratio of the excess events in $\MUis
\rightarrow \gamma K^+K^-$ 
with di-kaon invariant mass between $2-3\mass$ is
\begin{equation*}
{\cal B}(\MUis \rightarrow \gamma K^+K^-) =
(1.14\pm0.08\pm0.10)\times10^{-5},
\end{equation*}
where the first uncertainty is statistical and the second
is systematic. The sources of systematic 
uncertainty are 1\% from the number of
\Uis~decays, 2\% from the Monte Carlo simulation of the track
reconstruction, 3\% (8\%) from the Monte Carlo efficiency modeling of the
event requirements in the
pion (kaon) mode, and 1\% to 3\% from the uncertainty in
the angular distribution measurements. We also assign a 15\% systematic
uncertainty to the $f'_2(1525)$ branching fraction 
from possible interference between the
$f_2(1270)$ and
$f'_2(1525)$ resonances, and less than a 1\% systematic uncertainty from high-momentum
neutral pions faking photons in the decay
$\MUis \rightarrow \rho\pi$, based on the upper limit
in~\cite{brian}. Finally, we include the uncertainties in the  
$f_2(1270)$  and  $f_2^{'}(1525)$ hadronic branching
ratios~\cite{pdg} in the systematic uncertainty. For our less
significant signal candidates, the
branching fraction central values, along with their 
significances and 90\% 
confidence level upper limits, are
shown in Table~\ref{Table:brs}. 

\begin{table}[ht]
\begin{center}
\caption{Branching fraction central value (BF), its statistical
significance, and its 90\% confidence level upper limit (UL), for each
signal candidate with a significance $<5\sigma$. In
the branching fraction central value, the 
first uncertainty is statistical and the second is
systematic. The first three table entries are product branching
fractions.}\label{Table:brs}
\begin{tabular}{c|c|c|c}
\hline
\hline
Channel & BF $(10^{-5})$ & Significance & UL $(10^{-5})$\\
\hline 
\hline 
$\Upsilon$(1S) $\rightarrow  \gamma f_0(980) \rightarrow \pi^+ \pi^-$ & 
$1.8^{+0.8}_{-0.7} \pm0.1$ & $4.3\sigma$ & $ < 3$ \\
$\Upsilon$(1S) $\rightarrow  \gamma f_4(2050) \rightarrow \pi^+ \pi^-$ & $0.37\pm0.14\pm0.03$ & $2.6\sigma$ & $< 0.6$\\ 
\hline 
$\Upsilon$(1S) $\rightarrow \gamma f_0(1710) \rightarrow K^+ K^-$ & 
$0.38\pm0.16\pm0.04$ & $3.2\sigma$ & $<0.7$ \\
\hline 
$\Upsilon$(1S)$ \rightarrow \gamma p \bar{p} $, $2\mass$$< m_{p\bar{p}} <$$3\mass$ &
$0.41\pm0.08\pm0.10$ &  $4.8\sigma$ & $< 0.6$\\
\hline 
\hline
\end{tabular}
\end{center}
\end{table}

\section{Determination of Upper Limits for $f_J(2220)$ and $X(1860)$
Production and Decay}

\begin{figure}
\includegraphics*[width=3.25in]{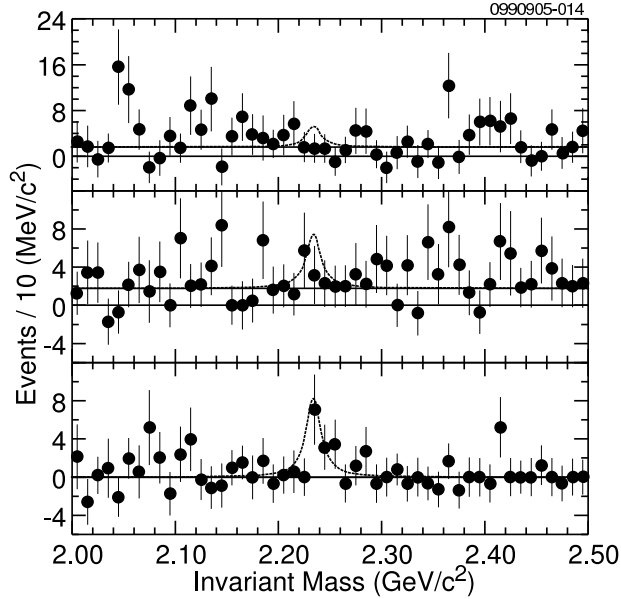}
\caption{Invariant mass of $\pi^+ \pi^-$ (top) $K^+K^-$ (middle) and
$p \bar{p}$ (bottom) from $\MUis
\rightarrow \gamma h^+h^-$ in the range 2.0 to 2.5 $\mass$.
Also shown are the results of fits to a $f_J(2220)$ Breit-Wigner
function (dashed) and a flat background (solid). The $f_J(2220)$ signal
line shown corresponds to the 90\% confidence level maximum yield 
obtained from the fit.
The flat background function is also used to measure the possible excesses
of events in the invariant mass region 2 to 3 $\mass$.}
\label{paper6.eps}
\end{figure}

To measure upper limits of the product branching ratio for the
decays \Uis$\rightarrow \gamma f_J(2220)$ with $f_J(2220) \rightarrow
h^+h^-$, we fit 
the $h^+h^-$ invariant mass plots, shown in 
Figure~\ref{paper6.eps}, using a Breit-Wigner with a peak mass and
width fixed at 2.234$\mass$ and 0.017$\mass$, 
respectively. These are the values from the possible $ f_J(2220)$
signal reported 
by the BES experiment~\cite{glueball}, which is considered a candidate for
a glueball. To model the
general excess of
events between 2.0 and 2.5$\mass$ we also use a flat background
function in the fit. 
Although
the highest bin in the $p \bar{p}$ plot is indeed in the region of the
$f_J(2220)$, the excess ($12\pm5$ events) is not significant, 
and there are no significant
signals anywhere in these three plots.
To find upper limits for $f_J(2220)\rightarrow h^+ h^-$ decays, we 
fix the area of the Breit-Wigner to different values, minimize the
chi-squared from the fit, and give that area a probability
proportional $e^{-\chi^2/2}$. These 
probability distributions are then used to obtain the following 90\%
confidence level upper limits on the product branching ratio for $f_J(2220)$
production and decay to each mode:  
\begin{equation*}
\begin{split}
&{\cal B}(\MUis \rightarrow \gamma f_J(2200))\times{\cal B}(f_J(2200) \rightarrow
\pi^+\pi^-) < 8\times10^{-7},
\\
&{\cal B}(\MUis \rightarrow \gamma f_J(2200))\times{\cal B}(f_J(2200) \rightarrow
K^+K^-) < 6\times10^{-7},
\\
&{\cal B}(\MUis \rightarrow \gamma f_J(2200))\times{\cal B}(f_J(2200) \rightarrow
p \bar{p}) < 11\times10^{-7}.
\end{split}
\end{equation*}
The systematic uncertainties on the branching ratios were added in
quadrature with the statistical errors in forming the above
limits. Using the $X(1860)$ parameters measured
in~\cite{ppbar}, and proceeding in a similar manner as described above, we
obtain: 
\begin{equation*} 
{\cal B}(\MUis\rightarrow \gamma X(1860))\times{\cal B}(X(1860)
\rightarrow p \bar{p}) < 5 \times10^{-7}, 
\end{equation*}
at the 90\% confidence level.  

\section{Summary and Conclusion}

We have confirmed CLEO's previous observation of the $f_2(1270)$ in
radiative \Uis~decays and made a new observation of the
$f_2^{'}(1525)$, obtaining factors of $0.07\pm0.01$ and
$0.08^{+0.04}_{-0.03}$ for the ratio of the \Uis~branching fraction
with respect to the one measured in $J/\psi$ radiative decays,
respectively. These values are larger than, but the same order of magnitude as,
the ratio of 0.04 expected from naive scaling arguments. The observed 
$f_2(1270)$ production is in agreement with the prediction in~\cite{nice}
and somewhat lower than the prediction in~\cite{ugg}.
In both of the measured modes we can confirm by
fits to  
the angular distributions of the photon and charged particles that the two
daughter hadrons are indeed produced by a spin-2 parent. We find 
that this parent is produced mostly with helicity 0, in good agreement with
the predictions in~\cite{nice,ugg}. 
No structure is
seen in the $p\bar{p}$ invariant-mass distribution. In particular, we do
not observe a 
near-threshold enhancement as in~\cite{ppbar}. Finally,  
stringent
limits have been put on the production of the glueball candidate $f_J(2220)$ 
in radiative \Uis~decays.
Glueball production is expected to be enhanced in \Uis~radiative
decays~\cite{cleof,pheno1,pheno2}, but we find that, within our experimental
sensitivity, known tensor 
meson states, believed to be composed only of quarks, dominate the
di-gluon spectrum. 

We gratefully acknowledge the effort of the CESR staff 
in providing us with excellent luminosity and running conditions.
This work was supported by the National Science Foundation
and the U.S. Department of Energy.


\begin{thebibliography}{99}


\bibitem{nrqcd}
  G.T.~Bodwin, E.~Braaten and G.P.~Lepage,
  Phys.\ Rev.\ D {\bf 51}, 1125 (1995).
\bibitem{ugg}
  S.~Fleming, C.~Lee and A.K.~Leibovich,
  Phys.\ Rev.\ D {\bf 71}, 074002 (2005).
\bibitem{nice}
  J.P.~Ma,
  Nucl.\ Phys.\ B {\bf 605}, 625 (2001).
\bibitem{russ1}
  V.N.~Baier and A.G. Grozin, 
  Sov.\ J.\ Nucl.\ Phys. {\bf 35}, 596 (1982).
\bibitem{russ2}
  V.N.~Baier and A.G.~Grozin, 
  Z.\ Phys.\ C {\bf 29}, 161 (1985).



\bibitem{pot1}
  T.~Barnes,
  Z.\ Phys.\ C {\bf 10}, 275 (1981).
\bibitem{pot2}
  J.M.~Cornwall and A.~Soni,
  Phys.\ Lett.\ B {\bf 120}, 431 (1983).
\bibitem{pot3}
  W.S.~Hou and G.G.~Wong,
  Phys.\ Rev.\ D {\bf 67}, 034003 (2003).
\bibitem{theory1}
 C.J. Morningstar and M.J. Peardon, 
 Phys. Rev. D {\bf 60}, 034509 (1999).
\bibitem{theory2}
 A. Vaccarino and D. Weingarten,
 Phys. Rev. D {\bf 60}, 114501 (1999).
\bibitem{lat1}
  C.~Liu,
  Commun.\ Theor.\ Phys.\  {\bf 35}, 288 (2001); Chin.\ Phys.\ Lett.\
{\bf 18}, 187 (2001).  
\bibitem{lat2}
  D.Q.~Liu, J.M.~Wu and Y.~Chen,
  High Energy Phys.\ Nucl.\ Phys.\  {\bf 26}, 222 (2002).



\bibitem{bag1}
  P.G.O.~Freund and Y.~Nambu,
  Phys.\ Rev.\ Lett.\  {\bf 34}, 1645 (1975).
\bibitem{bag2}
  R.L.~Jaffe and K.~Johnson,
  Phys.\ Lett.\ B {\bf 60}, 201 (1976).
\bibitem{bag3}
  T.~Barnes, F.E.~Close and S.~Monaghan,
  Nucl.\ Phys.\ B {\bf 198}, 380 (1982).
\bibitem{theory3}
 C.E. Carlson, T.H. Hansson, and C. Peterson,
 Phys. Rev. D {\bf 27}, 1556 (1983).
\bibitem{theory4}
 N. Isgur and J. Paton, 
 Phys. Rev. D {\bf 31}, 2910 (1985).
\bibitem{theory5}
  S.~Narison,
  Z.\ Phys.\ C {\bf 26}, 209 (1984); Nucl.\ Phys.\ B {\bf 509}, 312
(1998); Nucl.\ Phys.\ Proc.\ Suppl.\  {\bf 96}, 244 (2001). 


\bibitem{bs1}
  M.H.~Thomas, M.~Lust and H.J.~Mang,
  J.\ Phys.\ G {\bf 18}, 1125 (1992).
\bibitem{bs2}
  J.Y.~Cui, J.M.~Wu and H.Y.~Jin,
  Phys.\ Lett.\ B {\bf 424}, 381 (1998).
\bibitem{theory6}
S.J.~Brodsky, A.S.~Goldhaber and J.~Lee,
Phys.\ Rev.\ Lett.\  {\bf 91}, 112001 (2003).
\bibitem{cleof}
  X.G.~He, H.Y.~Jin and J.P.~Ma,
  Phys.\ Rev.\ D {\bf 66}, 074015 (2002).
\bibitem{maurizio}
 M.~Melis, F.~Murgia and J.~Parisi,
 Phys.\ Rev.\ D {\bf 70}, 034021, (2004).
\bibitem{sre}
  J.C.~Su and J.X.~Chen,
  Phys.\ Rev.\ D {\bf 69}, 076002 (2004).


\bibitem{expglu1}
  V.V.~Anisovich {\it et al.}  (Crystal Barrel Collaboration),
  Phys.\ Lett.\ B {\bf 323}, 233 (1994).
\bibitem{expglu2}
  V.V.~Anisovich, D.V.~Bugg, A.V.~Sarantsev and B.S.~Zou,
  Phys.\ Rev.\ D {\bf 50}, 1972 (1994).
\bibitem{expglu3}
  C.~Amsler {\it et al.}  (Crystal Barrel Collaboration),
  Phys.\ Lett.\ B {\bf 355}, 425 (1995).
\bibitem{experiment1}
 C.A. Meyer,
 ``Proceedings of the Workshop on Gluonic Excitations,'' 
 Newport News, Virginia 2003, AIP Conf. Proc. {\bf 698}, 554 (2004).
\bibitem{experiment2}
 K.K. Seth, 
 Nucl. Phys. B (Proc. Suppl.) {\bf 96}, 205 (2001).
\bibitem{experiment3}
 B.S.~ Zou, 
 Nucl. Phys. A{\bf 644}, 41c (1999).
\bibitem{experiment4}
  A.~Etkin {\it et al.},
  Phys.\ Lett.\ B {\bf 201}, 568 (1988); Phys.\ Lett.\ B {\bf 165}, 217 (1985).




\bibitem{noglu1}
  C.~Amsler and F.E.~Close,
  Phys.\ Lett.\ B {\bf 353}, 385 (1995).
\bibitem{noglu2}
  D.~Weingarten,
  Nucl.\ Phys.\ Proc.\ Suppl.\  {\bf 53}, 232 (1997).
\bibitem{noglu3}
  D.V.~Bugg, M.J.~Peardon and B.S.~Zou,
  Phys.\ Lett.\ B {\bf 486}, 49 (2000).
\bibitem{noglu4}
  J.~Sexton, A.~Vaccarino and D.~Weingarten,
  Phys.\ Rev.\ Lett.\  {\bf 75}, 4563 (1995).
\bibitem{noglu5}
  V.V.~Anisovich,
  Phys.\ Lett.\ B {\bf 364}, 195 (1995).
\bibitem{noglu6}
  F.~Giacosa, T.~Gutsche and A.~Faessler,
  Phys.\ Rev.\ C {\bf 71}, 025202 (2005).
\bibitem{noglu7}
  M.~Chanowitz,
  hep-ph/0506125.
\bibitem{noglu8}
  A.H.~Fariborz,
  Int.\ J.\ Mod.\ Phys.\ A {\bf 19}, 5417 (2004).
\bibitem{noglu9}
  A.V.~Anisovich, V.V.~Anisovich, Y.D.~Prokoshkin and A.V.~Sarantsev,
  Z.\ Phys.\ A {\bf 357}, 123 (1997);
  Nucl.\ Phys.\ Proc.\ Suppl.\ A {\bf 56}, 270 (1997).
\bibitem{noglu10}
  D.~Weingarten,
hep-ph/9607212.



\bibitem{tensor1}
  V.V.~Anisovich, M.A.~Matveev, J.~Nyiri and A.V.~Sarantsev,
  hep-ph/0506133;
  V.V.~Anisovich and A.V.~Sarantsev,
  hep-ph/0504106.
\bibitem{tensor2}
  Y.~D.~Prokoshkin {\it et al.} (GAMS Collaboration),
  Phys.\ Dokl.\  {\bf 40}, 495 (1995).

\bibitem{QWG}
  N.~Brambilla {\it et al.},
  hep-ph/0412158.

\bibitem{pdg}
 S. Eidelman {\it et al.} (Particle Data Group Collaboration), Phys. Lett. B
 {\bf 592}, 1 (2004).


\bibitem{pluto}
  G.~Alexander {\it et al.}  (Pluto Collaboration),
  Phys.\ Lett.\ B {\bf 76}, 652 (1978).
\bibitem{markII}
D.L.~Scharre, ``$10^{th}$ International Symposium on Lepton and Photon
Interactions at High Energy'', Bonn (1981).
\bibitem{crystal}
  C.~Edwards {\it et al.} (Crystal Ball Collaboration),
  Phys.\ Rev.\ D {\bf 25}, 3065 (1982).
\bibitem{dm2}
  J.E.~Augustin {\it et al.}  (DM2 Collaboration),
  Z.\ Phys.\ C {\bf 36}, 369 (1987).
\bibitem{markIII}
 R.M.~Baltrusaitis {\it et al.} (Mark III Collaboration),
 Phys. Rev. D {\bf 35}, 2077 (1987).
\bibitem{kkBes}
J.Z.~Bai {\it et al.}  (BES Collaboration),
Phys.\ Rev.\ D {\bf 68}, 052003 (2003).




\bibitem{very_old}
M.~Krammer,
Phys.\ Lett.\ B {\bf 74}, 361 (1978).
\bibitem{correction2}
  J.G.~Korner, J.H.~Kuhn, M.~Krammer and H.~Schneider,
  Nucl.\ Phys.\ B {\bf 229}, 115 (1983).






\bibitem{glueball}
 J.Z. Bai {\it et al.} (BES Collaboration),
 Phys. Rev. Lett. {\bf 76}, 3502 (1996).
\bibitem{markIIIglue}
  R.M.~Baltrusaitis {\it et al.}  (MARK-III Collaboration),
  Phys.\ Rev.\ Lett.\  {\bf 56}, 107 (1986).
\bibitem{dm2glue}
  J.E.~Augustin {\it et al.}  (DM2 Collaboration),
  Phys.\ Rev.\ Lett.\  {\bf 60}, 2238 (1988).





\bibitem{ppbar}
J.Z.~Bai {\it et al.} (BES Collaboration),
Phys.\ Rev.\ Lett.\  {\bf 91}, 022001 (2003).
\bibitem{pp1}
  A.~Sibirtsev, J.~Haidenbauer, S.~Krewald, U.G.~Meissner and A.W.~Thomas,
  Phys.\ Rev.\ D {\bf 71}, 054010 (2005).
\bibitem{pp2}
  B.~Kerbikov, A.~Stavinsky and V.~Fedotov,
  Phys.\ Rev.\ C {\bf 69}, 055205 (2004).
\bibitem{pp3}
  D.V.~Bugg,
  Phys.\ Lett.\ B {\bf 598}, 8 (2004).
\bibitem{pp4}
  C.S.~Gao and S.L.~Zhu,
  Commun.\ Theor.\ Phys.\  {\bf 42}, 844 (2004).
\bibitem{pp5}
  B.S.~Zou and H.C.~Chiang,
  Phys.\ Rev.\ D {\bf 69}, 034004 (2004).
\bibitem{pp6}
  B.~Loiseau and S.~Wycech,
  hep-ph/0501112.
\bibitem{pp7}
  G.J.~Ding and M.L.~Yan,
  Phys.\ Rev.\ C {\bf 72}, 015208 (2005).


\bibitem{cleo1}
 A. Anastassov {\it et al.} (CLEO Collaboration),
 Phys. Rev. Lett. {\bf 82}, 286 (1999).
\bibitem{cleo2}
 G. Masek {\it et al.} (CLEO Collaboration),
 Phys. Rev. D {\bf 65}, 072002 (2002).
\bibitem{LUIS}
 L. Breva-Newell, Ph.D. Thesis, University of Florida, hep-ex/0412075.
\bibitem{cleoiii}
Y. Kubota {\it et al.}, 
Nucl. Inst. Meth A {\bf 320} 66 (1992);
G. Viehhauser {\it et al.},
Nucl. Inst. Meth. A {\bf 462}, 146 (2001);
D.~Peterson {\it et al.}, Nucl. Inst. Meth. A {\bf 478}, 142
(2002); A. Warburton {\it et al.}, Nucl. Inst. and Meth. A {\bf 488},
451 
(2002);
M.~Artuso {\it et al.},
Nucl. Inst. Meth A {\bf 502} 91 (2003),
M.~Artuso {\it et al.},
physics/0506132.

  
\bibitem{nuis}
R.A.~Briere {\it et al.}  (CLEO Collaboration),
Phys.\ Rev.\ D {\bf 70}, 072001 (2004).
\bibitem{qq}
{\it QQ - The CLEO Event Generator,}  http://www.lns.cornell.edu/public/CLEO/soft/QQ (unpublished).
\bibitem{geant}
R. Brun {\it et al.}, GEANT 3.21, CERN Program Library
Long Writeup W5013 (1993), unpublished.
\bibitem{trigger}
M.A.~Selen, R.M.~Hans and M.J.~Haney,
IEEE Trans.\ Nucl.\ Sci.\  {\bf 48}, 562 (2001).
\bibitem{helicity}
 J.D. Richman 
 CALT-68-1148 (1984) (unpublished).



\bibitem{old_embarrasment}
  P.K.~Kabir and A.J.G.~Hey,
  Phys.\ Rev.\ D {\bf 13}, 3161 (1976).
\bibitem{embarrasment}
  M.~Ablikim {\it et al.}  (BES Collaboration),
  Phys.\ Rev.\ D {\bf 70}, 092004 (2004).






\bibitem{brian}
  S.A.~Dytman {\it et al.} (CLEO Collaboration),
  hep-ex/0307035.

\bibitem{pheno1}
  S.~Godfrey and J.~Napolitano,
  Rev.\ Mod.\ Phys.\  {\bf 71}, 1411 (1999).
\bibitem{pheno2}
  F.E.~Close, G.R.~Farrar and Z.~Li,
  Phys.\ Rev.\ D {\bf 55}, 5749 (1997).



\end{thebibliography}
\end{document}